\documentclass[conference]{IEEEtran}

\usepackage{colortbl}
\usepackage{url}

\usepackage{ifthen}
\newboolean{long_version}
\setboolean{long_version}{true}

\usepackage{multirow}

\ifCLASSINFOpdf
  \usepackage[pdftex]{graphicx}
   \graphicspath{{graphics/}}
\else
   \usepackage[dvips]{graphicx}
   \DeclareGraphicsExtensions{.eps,.ps}
\fi
\usepackage[tight,footnotesize]{subfigure}
\begin{document}
%
\title{Correlated Resource Models of Internet End Hosts}

\author{\IEEEauthorblockN{Eric M. Heien and Derrick Kondo}
\IEEEauthorblockA{INRIA\\
France\\
Email: \{eric.heien, derrick.kondo\}@inria.fr}
\and
\IEEEauthorblockN{David P. Anderson}
\IEEEauthorblockA{Space Sciences Laboratory\\
University of California, Berkeley, CA\\
Email: davea@ssl.berkeley.edu}
}

\maketitle

\begin{abstract}
  Understanding and modelling resources of Internet end
  hosts is essential for the design of desktop software and
  Internet-distributed applications.  In this paper we
  develop a correlated resource model of Internet end hosts
  based on real trace data taken from the SETI@home project.
  This data covers a 5-year period with statistics for 2.7
  million hosts.  The resource model is based on statistical
  analysis of host computational power, memory, and storage
  as well as how these resources change over time and the
  correlations between them.  We find that resources with
  few discrete values (core count, memory) are well modeled
  by exponential laws governing the change of relative
  resource quantities over time.  Resources with a
  continuous range of values are well modeled with either
  correlated normal distributions (processor speed for
  integer operations and floating point operations) or
  log-normal distributions (available disk space). We
  validate and show the utility of the models by applying
  them to a resource allocation problem for
  Internet-distributed applications, and demonstrate their value
  over other models.  We also make our trace data
  and tool for automatically generating realistic Internet
  end hosts publicly available.
\end{abstract}

%
\IEEEpeerreviewmaketitle

\section{Introduction}


While the Internet plays a vital role in society, relatively
little is known about Internet end hosts and in particular
their hardware resources.  Obtaining detailed data about
hardware resources of Internet hosts at a large-scale is
difficult.  The diversity of host ownership and privacy
concerns often preclude the collection of hardware
measurements across a large number of hosts.  Internet
safeguards such as firewalls make remote access to end hosts
almost impossible.  Also, ISPs are reluctant to collect or
release data about their end hosts.

Nevertheless, the characteristics and models of Internet end
hosts are essential for the design and implementation of any
desktop software or Internet-distributed application.  Such
software or applications include but are not limited to
operating systems, web browsers, peer-to-peer (P2P),
gaming, multi-media and word-processing applications.

Models are also needed for Internet-computing research.  For
instance, in works such as~\cite{issam_jpdc10, massimo_pcgrid08, lo_wave},
researchers developed algorithms for scheduling
or resource discovery for distributed applications run
across Internet hosts.  Assumptions had to be made about the
distribution of hardware resources of these Internet hosts,
and the performance of such algorithms are arguably tied to
the assumed distributions.  Realistic models of Internet
resources derived systematically from real-world data are
needed to quantify and understand the performance of these
algorithms under a range of scenarios.



Our goal in this study is to characterize and model
resources of Internet end hosts.  Our approach for data
collection is to use hardware statistics and measurements
retrieved by SETI@home.  SETI@home is one of the largest
volunteer computing projects in the world, aggregating
millions of volunteered hosts for distributed computation.
Using the SETI@home framework, we retrieved hardware data
over a 5 year period with statistics for 2.7 million hosts.

Our approach for modelling is to investigate statistically
the distribution, correlation, and evolution of resources.
Our main contributions are as follows:

\begin{enumerate}
\item We characterize and statistically model hardware
  resources of Internet hosts, including the number of
  cores, host memory, floating point/integer speed and disk
  space.  Our model captures the resource mixture across
  hosts and how it evolves over time.  Our model
  also captures the correlation of resources (for instance
  memory and number of cores) within individual hosts.

\item We evaluate the utility of our model and show its
  accuracy in the context of a resource allocation problem
  involving Internet distributed computing applications.

\item We make our data and tool for automated model
  generation publicly available.  Our model can be used to
  generate realistic sets of Internet hosts of today or
  tomorrow.  Our model can also be used to predict hardware
  trends.
\end{enumerate}

The paper is structured as follows.  In Section \ref{rel-work-sec} we discuss related work and how our contribution fits in.  We then discuss the application context for our model in Section \ref{app-context} and go over the data collection methodology in Section \ref{sec-data-collection}. We introduce details of the
model and describe how the resources are
modeled over time in Section \ref{sec-modelling}.
We validate the model using statistical techniques in Section
\ref{sec-res-predict} and show how it can be used to
generate realistic sets of hosts for simulations.  To demonstrate
the effectiveness of our model compared to other methods we
perform simulations in Section \ref{sec-model-sim}.
Finally, we offer discussion and future areas of work in Section \ref{conc-sec}.

\section{Related Work}
\label{rel-work-sec}

The branches of work related to this paper include Internet network modelling,
peer-to-peer (P2P) network modelling, desktop benchmarking,
and Grid resource modelling.

With respect to Internet network measurement and
modelling~\cite{floyd_ccr03, caida, powerlaw_sigcomm99},
previous studies tend to focus exclusively on the network of
end hosts, and not their hardware resources.  Several works
such as~\cite{neti, dimes,broadband_saroiu07} model
specifically residential networks, but omit hardware
measurements or models.  Also, the scale of those
measurements are relatively small on the order of thousands
of hosts monitored on the order of months (versus millions
of hosts on the order of years).  P2P research~\cite{saroiu_mmcn, chu_ITCom}
has focused primarily on application-level network traffic, topology, and its
dynamics.  Again, hardware measurements and models are missing.

For desktop benchmarking there are a handful of programs such as XBench~\cite{xbench},
PassMark~\cite{passmark} and LMBench~\cite{lmbench}.  However, these benchmarks are generally
designed for a particular operating system and set of tests - often oriented towards game graphics performance - 
making it difficult to compare across platforms.  These
benchmarks are also generally run only once on a system,
limiting their usefulness in predicting how total resource
composition changes over time.

Some previous works investigated modelling clusters or
computational Grids
\cite{Kee:2004p763,Sulistio:2008p7929,dinda_gridg}.  These
works differ from ours in terms of the resource focus of the
model, the host heterogeneity and the evolution and
correlation of resources over time.  Also, most Grid
resource models are based on data from many years ago and
may no longer be valid for present configurations.

The closest work described in~\cite{Anderson:2006p180} gives
a general characterization of Internet host resources.  However,
statistical models are not provided, and the evolution and
dynamics of Internet resources are not investigated.  Also,
certain hardware attributes (such as cores) are
not characterized or modeled due to the technology available
at that time.

\section{Application Context}
\label{app-context}

While there are an infinite range of host resources to
monitor and model, we select only those host properties that
are the most relevant for Internet distributed computing.
One class of Internet distributed computing is distributed
peer-to-peer (P2P) file
sharing~\cite{saroiu_mmcn,chu_ITCom,iptps_bhagwan}.  Another
important class is volunteer distributed computing.  As of
November 2010, volunteer computing provides over 7 PetaFLOPS
of computing power~\cite{Anderson:2002p327,Larson:2004p132}
for over 68 applications from a wide range of scientific
domains (including climate prediction, protein folding, and
gravitational physics). These projects have produced
hundreds of scientific result~\cite{boincpapers} published in the
world's most prestigious conferences and journals, such as
Science and Nature.  We use these types of application to
drive what we model.

\section{Data Collection Method}
\label{sec-data-collection}

The hosts in this study were measured using the BOINC
(Berkeley Open Infrastructure for Network Computing)
\cite{Anderson:2004p59} client software, and participated in
the SETI@home project \cite{Anderson:2002p327} between
January 1, 2006 and September 1, 2010.  We feel this data
set provides a reasonable approximation to the types of
hosts likely to be available for large scale Internet
computing applications.  The host model developed in this
paper uses the host data from January 1, 2006 to January 1,
2010.  We then validate this model by predicting the host
composition until September 1, 2010.

In BOINC projects, hosts perform work in a master-worker
style computing environment where the host is the worker and
the project server is the master.  Host resource
measurements occur every time the host contacts the server,
this allows the server to allocate the appropriate work for
the available host resources.  The host resource
measurements are recorded on the server and periodically
written to publicly available files.

\section{Modelling}
\label{sec-modelling}

In this section we discuss the model of host resources - how
it is defined and how we model the host resources and their
change over time.  In Section \ref{sec-host-overview} we
provide a general statistical overview of the hosts and how
the resources change over time.  Since two
resources may be correlated due to technological advancement
or user requirements, we begin the model building process by
examining correlation between resources in Section
\ref{sec-res-corr}.  In Sections \ref{sec-model-core}
through \ref{sec-model-disk} we perform detailed analysis of
each resource and build a predictive correlated model of
host cores, memory, computing speed and disk storage.
\ifthenelse{\boolean{long_version}}{ Finally, we briefly
  examine the characteristics of GPUs on hosts in Section
  \ref{sec-gpu-analysis}.  }{}

\subsection{Host Model}

\label{sec-host-model}

First we describe the model of hosts, including the
different resources in the model and how they were measured.

Given the application context described in
Section~\ref{app-context}, we consider hosts to have 5 key
resources:

\begin{itemize}
\item{\textbf{Processing Cores}: the number of primary processing cores.  This does not include GPU cores or other special purpose secondary processors.  For Windows machines this was measured by the {\tt GetSystemInfo} function, for Apple/Linux/Unix machines by the {\tt sysconf}, {\tt sysctl} or similar functions.}
\item{\textbf{Integer computing speed}: the speed of a processing core as measured by the Dhrystone \cite{Weicker:1984p7562} 2.1 benchmark in C.}
\item{\textbf{Floating point computing speed}: the speed of a core as measured by the 1997 Whetstone benchmark in C \cite{Curnow76asynthetic}.}
\item{\textbf{Volatile Memory}: Random access memory used by the processors during computation.  For Windows machines this was measured by the {\tt GlobalMemoryStatusEx} function, for Apple/Linux/Unix machines by the {\tt Gestalt}, {\tt sysconf} and {\tt getsysinfo} functions.}
\item{\textbf{Non-volatile storage}: unused space in long term storage including hard disk drives.  This does not necessarily include all storage devices attached to a host, only those accessible to the BOINC client.  For Windows machines this was measured by the {\tt GetDiskFreeSpaceEx} function, for Apple/Linux/Unix machines by the {\tt statfs} or {\tt statvfs} functions.}
\end{itemize}

Although Whetstone and Dhrystone have various shortcomings, we feel their use is acceptable as an approximate measure of host computational ability.  In the official BOINC distribution these benchmarks were compiled using the -O2 flag for the UNIX version, the -Os flag for the Mac version using XCode and the /O2 /Ob1 flags for Windows version using Visual Studio.  Users can compile their own version of the benchmark code, however, very few choose to do so and therefore the executed measurement code can be viewed as being mostly homogeneous.  The benchmarks are executed on all available cores simultaneously and the average speed is taken.  Therefore, shared resources on multicore machines may adversely affect processor performance results.

Hosts may also have GPU coprocessors which can be used for GPGPU computing.  BOINC did not start recording GPU statistics until September 2009 when 12.7\% of active hosts reported having GPUs.  By September 2010, 23.8\% of active hosts reported having GPUs.
\ifthenelse{\boolean{long_version}}{
We feel one year of sampling provides insufficient data to include GPU characteristics in our model, however, we include a brief analysis of host GPUs in Section \ref{sec-gpu-analysis}.
}{However, we feel one year of sampling provides insufficient data to include GPU characteristics in our model.}

For the purposes of measuring host characteristics, a host is considered to be active at a time $T$ if the host first connected to the server before time $T$ and the most recent connection to the server is after time $T$.  Because we care about the aggregate statistics of hosts, we did not consider host availability at a detailed level.  For more fine-grained analysis of host availability see \cite{Javadi:2009p7323,Nurmi:2005p333}.

\subsection{Host Overview}
\label{sec-host-overview}

\begin{figure}[!t]
\centering
\includegraphics[width=0.47\textwidth]{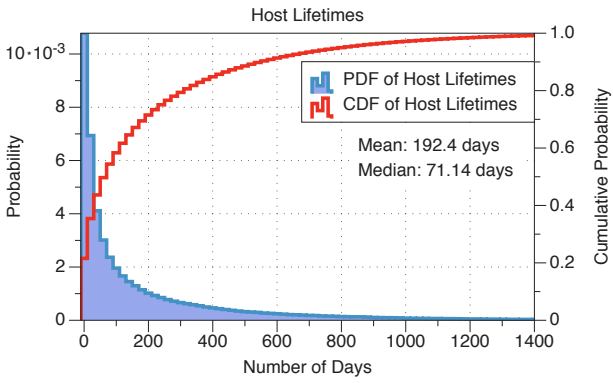}
\caption{Distribution of host lifetimes.}
\label{fig-host-lifetime}
\end{figure}

First we present an overview of the active hosts and their resources.  Figure \ref{fig-host-lifetime} shows a probability density function (PDF) and cumulative distribution function (CDF) of host lifetimes, where the lifetime is defined as the time between the first and last connection of the host to the server.  To avoid biasing the distribution towards short host lifetimes, this does not include hosts which connected after July 1, 2010.  Using a maximum likelihood of fit estimation we find the host lifetime distribution fits well to a Weibull distribution with parameters $k=0.58, \lambda=135$, which indicates that hosts have a decreasing dropout rate.

Some host data values may be questionable due to storage/transmission errors or modification of the client resource checking function.  In this paper, we discard hosts which report more than 128 cores, $10^5$ Whetstone MIPs, $10^5$ Dhrystone MIPs, $10^2$ GB memory or $10^4$ GB available disk space.  Based on these criteria we discard 3361 hosts (0.12\% of total).

\begin{figure}[!t]
\centering
\includegraphics[width=0.47\textwidth]{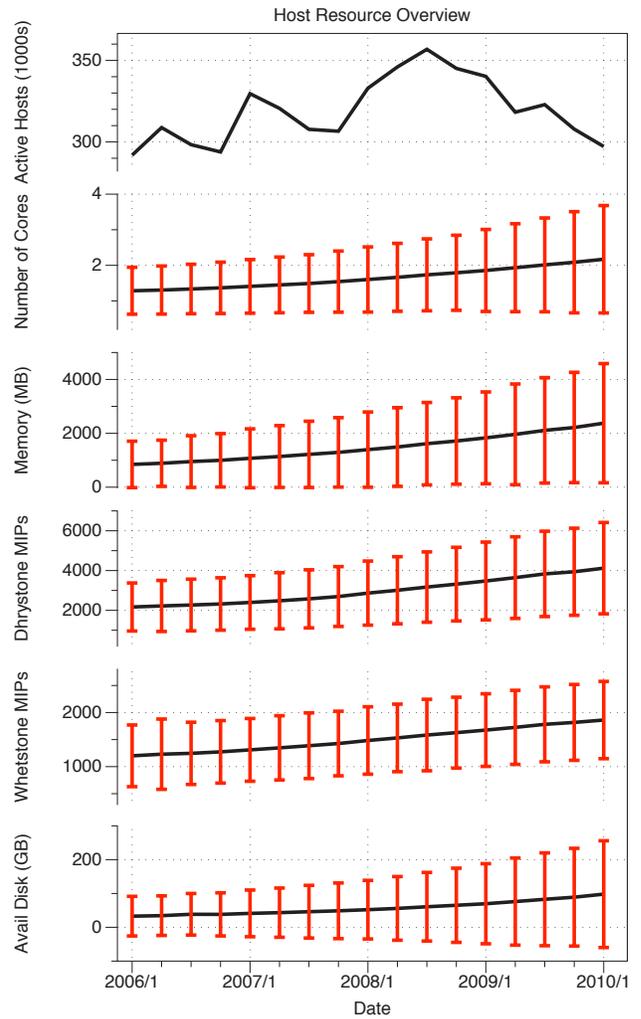}
\caption{Overview of host statistics, including number of active hosts and averages/standard deviations of number of cores, memory, per core integer and floating point speed and available disk space.}
\label{fig-res-overview}
\end{figure}

Figure \ref{fig-res-overview} shows the number of active hosts, and the mean and standard deviation of resources (cores, memory, computing speed and storage) over a 4 year period.  The mean of resource values is indicated by a black line, the standard deviation by red error bars.  The number of active hosts fluctuates between roughly 300,000 and 350,000.

This figure shows the changes in average host resources over 4 years.  From 2006 to 2010, the average number of cores in a host rose from 1.28 to 2.17 (70\% increase), the average memory rose from 846 MB to 2376 MB (181\% increase), the floating point performance rose from 1200 MIPS to 1861 MIPS (55\% increase), the integer performance rose from 2168 MIPS to 4120 MIPS (90\% increase) and the average available disk space rose from 32.9 GB to 98.0 GB (198\% increase).  The standard deviation of all resources  increased over time.  However, the increases in mean resource value are somewhat less than would be expected from Moore's law.

\begin{figure}[!t]
\centering
\includegraphics[width=0.47\textwidth]{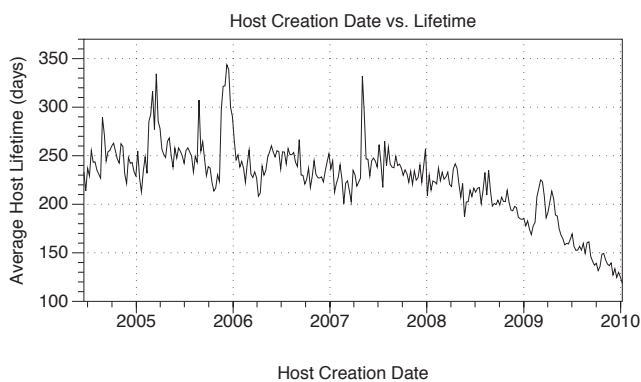}
\caption{Host creation date vs. average lifetime.}
\label{fig-creation-vs-lifetime}
\end{figure}

After closer investigation, we discovered this to be related to host lifetime.  As shown in Figure \ref{fig-creation-vs-lifetime}, there is a negative correlation between recently created hosts and host lifetime.  This means that more up to date hosts will tend to be underrepresented in the model.  We found similar patterns in speed and memory where hosts with better resources tended to have a shorter lifetime, though the reasons for this are unclear.

\ifthenelse{\boolean{long_version}}{

\begin{table}
\caption{Host processors over time (\% of total).}
\centering
\scriptsize
\begin{tabular}{|c|c|c|c|c|c|}
\hline
& 2006 & 2007 & 2008 & 2009 & 2010 \\
\hline
PowerPC G3/G4/G5 & 5.1 & 6.5 & 4.7 & 3.5 & 2.7 \\
\hline
\hline
Athlon XP & 12.3 & 9.0 & 6.2 & 4.0 & 2.5 \\
\hline
Athlon 64 & 6.5 & 9.5 & 11.4 & 11.6 & 10.2 \\
\hline
Other AMD & 8.3 & 8.2 & 7.8 & 7.9 & 9.5 \\
\hline
\hline
Pentium 4 & 36.8 & 33.0 & 27.2 & 20.7 & 15.5 \\
\hline
Pentium M & 5.4 & 5.5 & 4.3 & 3.1 & 2.1 \\
\hline
Pentium D & 0.7 & 3.0 & 4.2 & 3.9 & 3.1 \\
\hline
Other Pentium & 4.1 & 2.6 & 2.1 & 3.3 & 5.2 \\
\hline
\hline
Intel Core 2 & 0.9 & 3.3 & 13.2 & 24.8 & 32.0 \\
\hline
Intel Celeron & 5.6 & 6.4 & 6.3 & 5.9 & 4.9 \\
\hline
Intel Xeon & 2.1 & 2.8 & 3.3 & 3.9 & 4.3 \\
\hline
\hline
Other x86 & 9.9 & 7.7 & 7.6 & 6.1 & 5.1 \\
\hline
Other & 2.3 & 2.6 & 1.6 & 1.3 & 2.9 \\
\hline
\end{tabular}
\label{proc-type-table}
\end{table}

We also examine the composition of processors among the hosts and how it has changed over time.  Because availability and performance of new processor models cannot be predicted far in the future, we do not include processor information in our model.  There is also a significant range of speeds and capabilities even within a single processor family, making it difficult to predict the effect on a particular application. 

Table \ref{proc-type-table} shows the change in processor composition as a percent of total over the data sample period.  Several things are apparent from this table.  First, the Pentium 4 and similar Pentium processors processor were dominant in 2006 comprising over a third of processors, but by 2010 fell significantly to comprise only 15\% of processors.  Pentium 4 processors stopped shipping in 2008, so we expect the numbers to fall further as the processors fail over time.  In place of the Pentium, the Intel Core 2 (started shipping in 2006) went from zero to nearly a third of available processors.  The Intel Core 2 will likely stop shipping by 2011 so we expect the share to fall in the near future.

\begin{table}
\caption{Host OS over time (\% of total).}
\centering
\scriptsize
\begin{tabular}{|c|c|c|c|c|c|}
\hline
& 2006 & 2007 & 2008 & 2009 & 2010 \\
\hline
Windows XP & 69.8 & 71.5 & 68.6 & 62.5 & 52.9 \\
\hline
Windows Vista & 0 & 0 & 6.7 & 14.0 & 15.9 \\
\hline
Windows 7 & 0 & 0 & 0 & 0 & 9.2 \\
\hline
Windows 2000 & 12.9 & 8.5 & 5.5 & 3.4 & 2.0 \\
\hline
Other Windows & 6.3 & 6.1 & 4.8 & 4.8 & 3.4 \\
\hline
Mac OS X & 5.4 & 7.8 & 7.9 & 8.5 & 9.0 \\
\hline
Linux & 5.1 & 5.7 & 6.0 & 6.4 & 7.3 \\
\hline
Other & 0.4 & 0.4 & 0.4 & 0.3 & 0.3 \\
\hline
\end{tabular}
\label{os-type-table}
\end{table}

Table \ref{os-type-table} shows the change in host operating system over the sample period.  During this period, hosts using Windows XP drop from roughly 70\% to 50\%, while Windows Vista and Windows 7 increase from 0\% to roughly 25\%.  The remainder of hosts use a mix of other Windows systems (5-20\%) or Mac OS X or Linux (10-15\%).  These results indicate that although Windows is still the most common operating system, the share of Mac and Linux is steadily growing.
}{}

\subsection{Resource Correlations}
\label{sec-res-corr}

\begin{table}
\caption{Correlation coefficients between host measurements.}
\centering
\scriptsize
\begin{tabular}{|c|c|c|c|c|c|c|}
\hline
& Cores & Memory & Mem/Core & Whet & Dhry & Disk \\
\hline
Cores & 1.00 & \cellcolor[gray]{.7} \textbf{0.606} & -0.010 & 0.161 & 0.130 & 0.089 \\
\hline
Memory & \cellcolor[gray]{.7} \textbf{0.606} & 1.00 &  \cellcolor[gray]{.7} \textbf{0.627} & \cellcolor[gray]{.9} \textbf{ 0.230} & \cellcolor[gray]{.9} \textbf{ 0.271} & 0.114 \\
\hline
Mem/Core & -0.010 & \cellcolor[gray]{.7} \textbf{ 0.627} & 1.00 & \cellcolor[gray]{.9} \textbf{ 0.250} & \cellcolor[gray]{.9} \textbf{ 0.306 }& 0.065 \\
\hline
Whet & 0.161 & \cellcolor[gray]{.9} \textbf{ 0.230 }&  \cellcolor[gray]{.9} \textbf{ 0.250 }& 1.00 & \cellcolor[gray]{.7} \textbf{ 0.639 }& -0.016 \\
\hline
Dhry & 0.130 & \cellcolor[gray]{.9} \textbf{ 0.271} & \cellcolor[gray]{.9} \textbf{ 0.306} & \cellcolor[gray]{.7} \textbf{ 0.639} & 1.00 & -0.004 \\
\hline
Disk & 0.089 & 0.114 & 0.065 & -0.016 & -0.004 & 1.00 \\
\hline
\end{tabular}
\label{corr-table}
\end{table}

To guide us in creating the model of host resources, we first examine the correlations between different resources.  All resources will tend to improve together as technology advances over time.  Also, users will tend to purchase systems with correlated resource characteristics, for example, a system with many cutting edge cores will also tend to have a greater amount of memory.  Therefore our model should include these correlations to realistically capture the characteristics of hosts.

Visual inspection of the data showed a linear correlation between certain resources.  Table \ref{corr-table} shows the normalized coefficient of correlation (often called the Pearson correlation coefficient) for host resources, with table entry X, Y showing the r-value for the correlation between resources X and Y.  This table includes the resource ``per-core-memory'' (defined as amount of memory divided by number of cores) since this will be useful in generating a model of memory.

Several things are immediately apparent from this analysis.  First, the number of cores and memory of the host is well correlated ($r > 0.6$), though the amount of memory per core is not well correlated to the number of cores.  Also, the number of cores is poorly correlated with the integer and floating point performance of each core.  This may be related to the shared use of memory and bus during the benchmark routines.

Performance of integer and floating point benchmarks are also well correlated with each other ($r > 0.6$).  This is due to advances in processor technology which tend to improve both floating point and integer performance.  The best correlation between benchmark performance and other resources is that with memory ($r \approx 0.3$) rather than cores.

One somewhat surprising finding is that available disk space is not well correlated with any other metric, indicating that disk space may be modeled by an independent random distribution.  This is likely because disk usage is heavily dependent on the individual behavior of each user.  We also found that the fraction of total disk which is available is well represented by a uniform random distribution.

This analysis indicates that hosts in the generative model should have similar correlations between resources.  For example, a host with more cores should tend to have more memory, which will have some correlation with both the integer and floating point performance of the cores.

\subsection{Modelling Multicore}
\label{sec-model-core}

\begin{figure}[!t]
\centering
\includegraphics[width=0.47\textwidth]{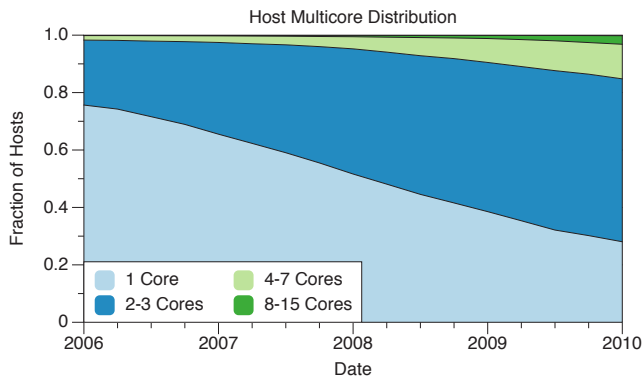}
\caption{Number of hosts and cores per host.}
\label{fig-ncpu-graph}
\end{figure}

\begin{figure}[!t]
\centering
\includegraphics[width=0.47\textwidth]{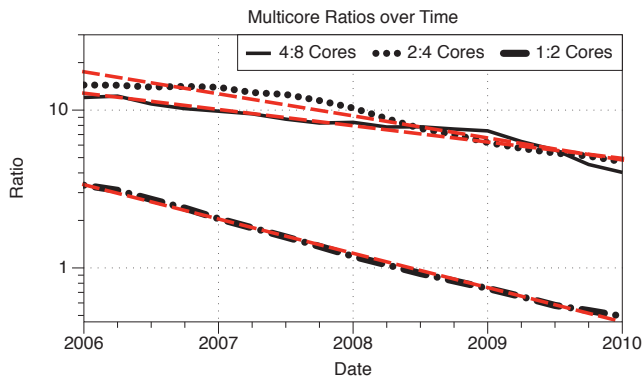}
\caption{Ratios of hosts with varying core numbers.  These are well fit by the function $a e^{b (year-2006)}$ (shown in red). Table \ref{core-ratio-val-table} has the $a$ and $b$ values.}
\label{fig-ratio-graph}
\end{figure}

\begin{table}
\caption{Core ratio model values.}
\centering
\begin{tabular}{|c|c|c|c|}
\hline
& $a$ & $b$ & $r$ \\
\hline
1:2 Core Ratio & 3.369 & -0.5004 & -0.9984 \\
\hline
2:4 Core Ratio & 17.49 & -0.3217 & -0.9730 \\
\hline
4:8 Core Ratio & 12.8 & -0.2377 & -0.9557 \\
\hline
\end{tabular}
\label{core-ratio-val-table}
\end{table}

In recent years, due to power and heat dissipation concerns, processor manufacturers have started increasing the number of cores on a processor rather than exclusively increasing the speed of the individual cores.  This trend is seen in Figure \ref{fig-ncpu-graph}, which shows the fraction of hosts with different numbers of cores over time.  In 2006, the ratio of 1 core machines to 2 core machines was 3.3 to 1, however, by 2010 the ratio inverted to 1 to 2.5 and 18\% of hosts had more than 4 cores.  There were not enough hosts in the data set with 16 or more cores for us to make a reasonable model of these machines.

Since the number of cores on a host is a discrete value, we are limited in the types of probability distributions we can use.  For the model of multicore on a host, we use a discrete probability distribution where the number of cores must be a power of 2.  Although there are systems available with non-power-of-two core counts, we ignore them since they comprise less than 0.3\% of hosts in our data set.   As processors with more cores are introduced to the marketplace, their number will increase relative to processors with fewer cores then decrease relative to processors with even more cores.  To model this, we examine the history of the ratio of 1, 2, 4 and 8 core hosts to each other since 2006.

Figure \ref{fig-ratio-graph} shows a logarithmic plot of the core ratios from 2006-2010.  The black lines show the actual ratios from the data set and the red dashed lines show the best fit.  For example, in 2006 there were roughly 14.4 2-core hosts for every 4-core host, but by 2010 this ratio had dropped to 4.7 2-core hosts for every 4-core host.  We found that the relative fractions of each of these is well modeled using an exponential function $a e^{b (year-2006)}$.  The values of $a$ and $b$ which best fit the data are shown in Table \ref{core-ratio-val-table} along with the correlation coefficient $r$.  In all cases the fitted curve has a very good match with the data.  Therefore, we can model the number of cores in a host as a ratio governed by an exponential function.

\subsection{Modelling Memory}
\label{sec-model-mem}
The available memory per host is also increasing over time as shown in Figure \ref{fig-res-overview}.  However, the analysis in Table \ref{corr-table} indicates a strong correlation ($r > 0.6$) between the number of cores and amount of memory.  Rather than trying to model host memory as a function of the cores, we instead model per-core-memory and multiply the results by the number of cores.
\ifthenelse{\boolean{long_version}}{
This makes intuitive sense - a host with 512 MB of RAM is more likely to have 1 core rather than 8 cores (which would be only 64 MB of RAM per core).  This is also}{This is}
 supported by the correlation analysis in Section \ref{sec-res-corr}, which showed that although the total memory is correlated with the number of cores, the amount of per-core-memory has nearly zero correlation and can therefore be generated independently of the number of cores.

\begin{figure}[!t]
\centering
\includegraphics[width=0.47\textwidth]{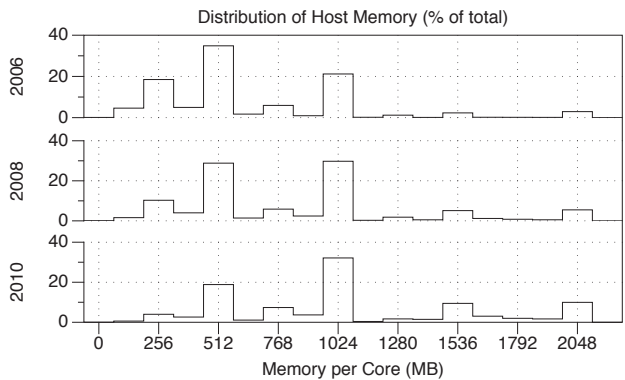}
\caption{Percent of hosts with varying per-core-memory in different years.}
\label{fig-mem-time-hist}
\end{figure}

First we examine the per-core-memory and how it changes over time.  Figure \ref{fig-mem-time-hist} shows distributions of per-core-memory at three points in time.  This figure shows a clear trend of per-core-memory increasing over time.  The fraction of hosts with 256MB or less per core drops from 19\% to 4\% of the total from 2006 to 2010, while the fraction of hosts with 1024MB per core rises from 21\% to 32\% and hosts with 2048MB per core rise from 2\% to 10\%.  Over 80\% of the per-core-memory values are in the set of (256, 512, 768, 1024, 1536, 2048) MB.  To simplify the model, we use these values to calculate the amount of memory on a host.

\begin{figure}[!t]
\centering
\includegraphics[width=0.47\textwidth]{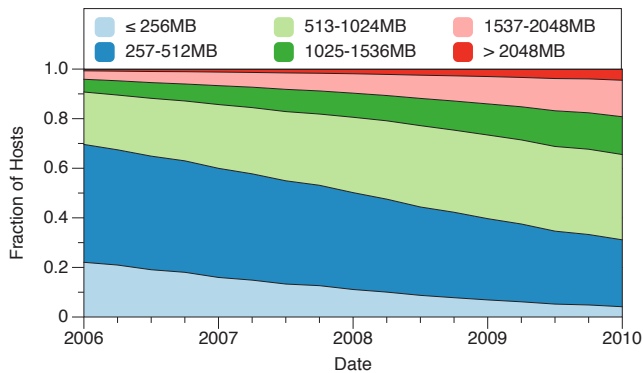}
\caption{Fractions of hosts with different per core memory.}
\label{fig-per-cpumem-frac-graph}
\end{figure}

Figure \ref{fig-per-cpumem-frac-graph} shows the fraction of hosts with different amounts of memory per core and how this changes over time.  Similar to multicore counts, we find that the ratios of host per-core-memory are best modeled by the exponential growth law $a e^{b (year-2006)}$.  The values for these ratios and their change over time is given in Table \ref{mem-ratio-val-table}.  The correlation coefficient $r$ indicates the values match the data very well.
\ifthenelse{\boolean{long_version}}{
It is worth noting that we discard some intermediate per-core-memory values (e.g. 1280MB, 1792MB, etc).  The accuracy of the model could therefore be improved by including these values, though at a cost of increased complexity.
}{}

\begin{table}
\caption{Per-core-memory ratio model values.}
\centering
\begin{tabular}{|c|c|c|c|}
\hline
& $a$ & $b$ & $r$ \\
\hline
256MB:512MB Ratio & 0.5829 & -0.2517 & -0.9984 \\
\hline
512MB:768MB Ratio & 4.89 & -0.1292 & -0.9748 \\
\hline
768MB:1GB Ratio & 0.3821 & -0.1709 & -0.9801 \\
\hline
1GB:1.5GB Ratio & 3.98 & -0.1367 & -0.9833 \\
\hline
1.5GB:2GB Ratio & 1.51 & -0.0925 & -0.9897 \\
\hline
2GB:4GB Ratio & 4.951 & -0.1008 & -0.9880 \\
\hline
\end{tabular}
\label{mem-ratio-val-table}
\end{table}

\subsection{Modelling Processor Speed}
\label{sec-model-speed}

\begin{figure}[!t]
\centering
\includegraphics[width=0.47\textwidth]{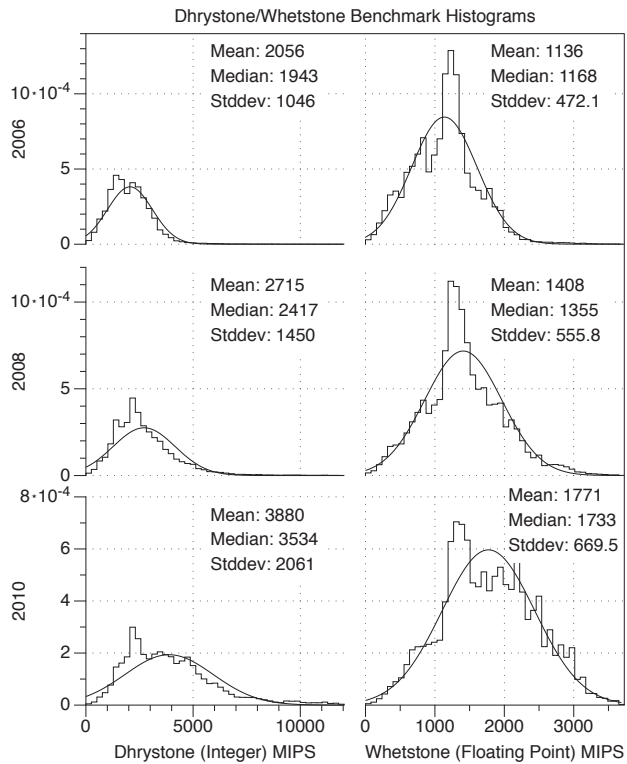}
\caption{Histograms of benchmark performance over time.}
\label{fig-benchmark-hist}
\end{figure}

Next we develop a model for host computational speed in
terms of Dhrystone and Whetstone benchmark performance.
Figure \ref{fig-benchmark-hist} shows histograms of the
Dhrystone and Whetstone MIPS performance at three times in
the data set.  First, we notice that the mean and standard
deviation of both measurements are increasing over time,
following the results we showed in Figure
\ref{fig-res-overview}.  To predict the mean and variance of
each benchmark we use a simple fitting on our data set.  We
found these values to be best fit by an exponential function
of the form $a e^{b (year-2006)}$ with $a$ and $b$ values
given in Table \ref{benchmark-ratio-val-table}.

To test the best fitting distribution for processor speeds
we used the Kolmogorov-Smirnov test.  This test is sensitive
to slight discrepancies in large data sets, so to calculate
p-values we took the average p-value of 100 KS tests each
using a randomly selected subset of 50 values. This
subsamping method is a standard method also used
in~\cite{Javadi:2009p7323,Nurmi:2005p333}.  We compared
our data to 7 distributions - normal, log-normal, exponential,
Weibull, Pareto, gamma and log-gamma. The results of
this show that the normal distribution fits the Whetstone
and Dhrystone data best with average p-values ranging from
0.19 to 0.43 at different times in the data.
\ifthenelse{\boolean{long_version}}{
Due to the spike around the middle of the distribution this is not a
perfect match, but we feel it is a reasonable model for
processor speed.}{}

However, we cannot simply choose the speeds from two normal
distributions since there is a strong correlation ($r >
0.6$) between the benchmarks and a slight correlation
($r \approx 0.3$) with memory.  To properly capture these correlations,
we create correlated statistics using a common method
involving the Cholesky decomposition.  We first take a
matrix $R$ of the correlation coefficients between
per-core-memory, Dhrystone and Whetstone performance from
Table \ref{corr-table}.

\[
R = \left[ {\begin{array}{ccc}
1 & 0.250 & 0.306 \\
0.250 & 1 & 0.639 \\
0.306 & 0.639 & 1 \\
\end{array} } \right]
\]

We apply the Cholesky decomposition to get matrix $U$.
\[
U = \left[ {\begin{array}{ccc}
1 & 0 & 0 \\
0.250 & 0.968 & 0 \\
0.306 & 0.581 & 0.754 \\
\end{array} } \right]
\]

We take a vector $V$ of three values randomly selected from a normal distribution with mean 0 and standard deviation 1.  $V_C = V U$ gives a vector of three values correlated by the values in $R$.  $V_C[1]$ is converted from a normal distribution to a uniform distribution and used to select the per-core-memory, $V_C[2]$ and $V_C[3]$ are renormalized to the predicted mean and variance for the Whetstone and Dhrystone benchmarks, respectively.  Using this method we are able to generate hosts with similar resource correlations as in the actual data.

\begin{table}
\caption{Benchmark and disk space prediction law values.}
\centering
\begin{tabular}{|c|c|c|c|}
\hline
& $a$ & $b$ & $r$ \\
\hline
Dhrystone Mean (MIPS) & 2064 & 0.1709 & 0.9946 \\
\hline
Dhrystone Variance & 1.379e6 & 0.3313 & 0.9937 \\
\hline
Whetstone Mean (MIPS) & 1179 & 0.1157 & 0.9981 \\
\hline
Whetstone Variance & 3.237e5 & 0.1057 & 0.8795 \\
\hline
Disk Space Mean (GB) & 31.59 & 0.2691 & 0.9955 \\
\hline
Disk Space Variance & 2890 & 0.5224 & 0.9954 \\
\hline
\end{tabular}
\label{benchmark-ratio-val-table}
\end{table}

\subsection{Modelling Available Disk Space}
\label{sec-model-disk}

Finally we develop the model for available disk space on a host.  As shown in Section \ref{sec-res-corr}, there is almost no correlation between available disk space and other resource metrics.  Because of this, we can safely generate a model of available disk space independent of the other resources.

Also, it is worth noting why we chose to model available disk space rather than total disk space.  The main reasons are: 1) total disk space is also uncorrelated with any other resource metric so we don't lose model accuracy, 2) the distribution of total disk space is highly irregular and difficult to model, 3) applications using Internet computing resources will generally be restricted by available disk space rather than total space.

\begin{figure}
\centering
\subfigure[Available disk space in 2006.]{
\includegraphics[width=0.46\textwidth]{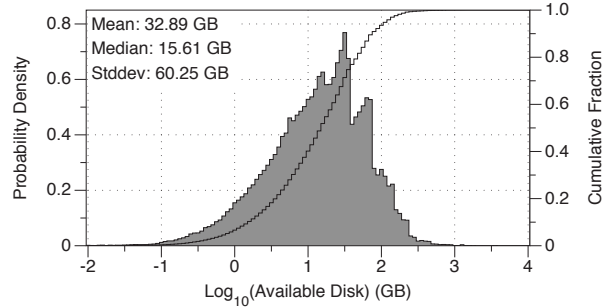}
\label{fig-disk-hist-2006}
}
\subfigure[Available disk space in 2008.]{
\includegraphics[width=0.46\textwidth]{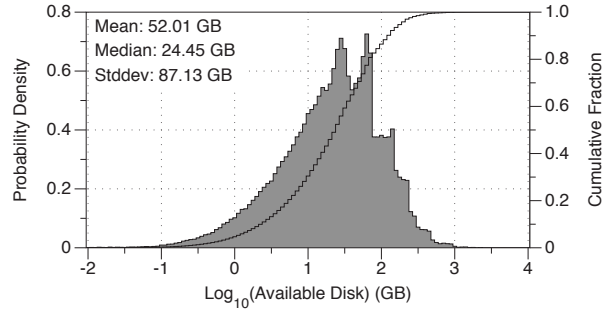}
\label{fig-disk-hist-2008}
}
\subfigure[Available disk space in 2010.]{
\includegraphics[width=0.46\textwidth]{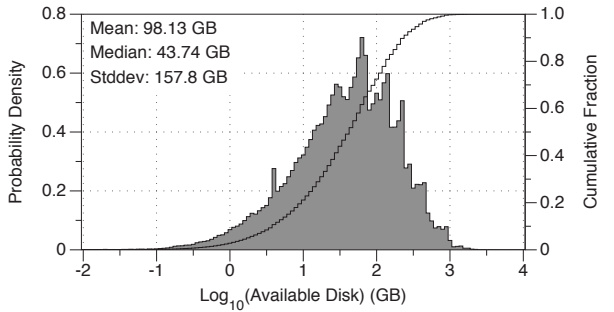}
\label{fig-disk-hist-2010}
}
\label{fig-disk-hist-all}
\caption{Histograms of available disk space over time.}
\end{figure}

Figures \ref{fig-disk-hist-2006}, \ref{fig-disk-hist-2008}
and \ref{fig-disk-hist-2010} show the probability density and cumulative distribution functions of the logarithm of available disk space on active hosts at three times.
\ifthenelse{\boolean{long_version}}{
The left sides of these distributions are smooth and fit well to a normal distribution.  The right side is somewhat less smooth with several spikes but still appears to fit reasonably well with a normal distribution.}{}
To test the best fitting distribution for disk space we again use the Kolmogorov-Smirnov test with the 7 distributions and average p-value.  The results show that the log-normal distribution best fits the data at different times with p-values ranging from 0.43 to 0.51.  Therefore we model available disk space as an independent log-normal distribution with mean and variance calculated using the exponential law with values from Table \ref{benchmark-ratio-val-table}.

\ifthenelse{\boolean{long_version}}{

\subsection{GPU Analysis}
\label{sec-gpu-analysis}

In recent years, GPU (graphics processing unit) based
computing has become popular and many computers include one
or more GPUs.  BOINC did not start recording GPU resource
information until September 2009, so we feel there is
insufficient data to include GPU resources in our model.
However, for completeness, we include a brief analysis of GPU resources in this section.

\begin{table}
\caption{Percent of GPU types among GPU equipped hosts.}
\centering
\begin{tabular}{|c|c|c|}
\hline
& Sep. 2009 & Sep. 2010 \\
\hline
GeForce & 82.5\% & 63.6\% \\
\hline
Radeon & 12.2\% & 31.5\% \\
\hline
Quadro & 4.7\% & 4.0\% \\
\hline
Other & 0.6\% & 0.8\% \\
\hline
\end{tabular}
\label{table-gpu-type}
\end{table}

Table \ref{table-gpu-type} shows a breakdown of the active hosts reporting GPUs based on the type of GPU they reported.  This breakdown is only among the 12.7\% (Sep. 2009) and 23.8\% (Sep. 2010) of hosts which reported having a GPU.

\begin{figure}[!t]
\centering
\includegraphics[width=0.46\textwidth]{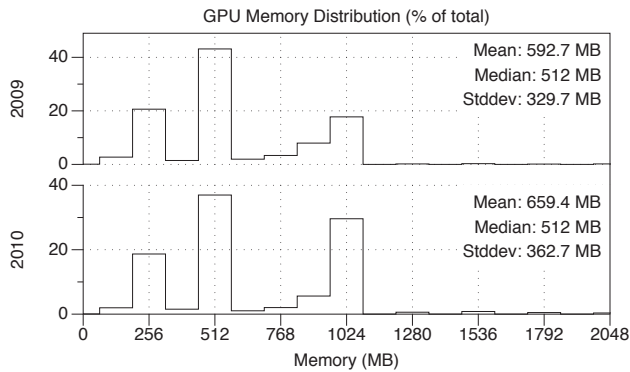}
\caption{GPU memory distribution at two times.}
\label{fig-gpu-memory}
\end{figure}

Figure \ref{fig-gpu-memory} shows the distribution of memory in GPUs from September 2009 and September 2010.  Between these dates, the average amount of GPU memory increased by 11\% from 592.7 MB to 659.4 MB.  There was a jump of GPUs with 1GB or more of memory from 19\% to 31\% of total.  However, these rises are significantly slower than the rate of increase in total host memory.  In addition, hosts with more than 1GB of GPU memory still comprise less than 2\% of GPU equipped hosts (0.5\% of all hosts), indicating memory bound applications may not be suitable for Internet end host GPUs in the near future.
}{}

\section{Model Validation and Prediction}
\label{sec-res-predict}

\begin{figure}[!t]
\centering
\includegraphics[width=0.47\textwidth]{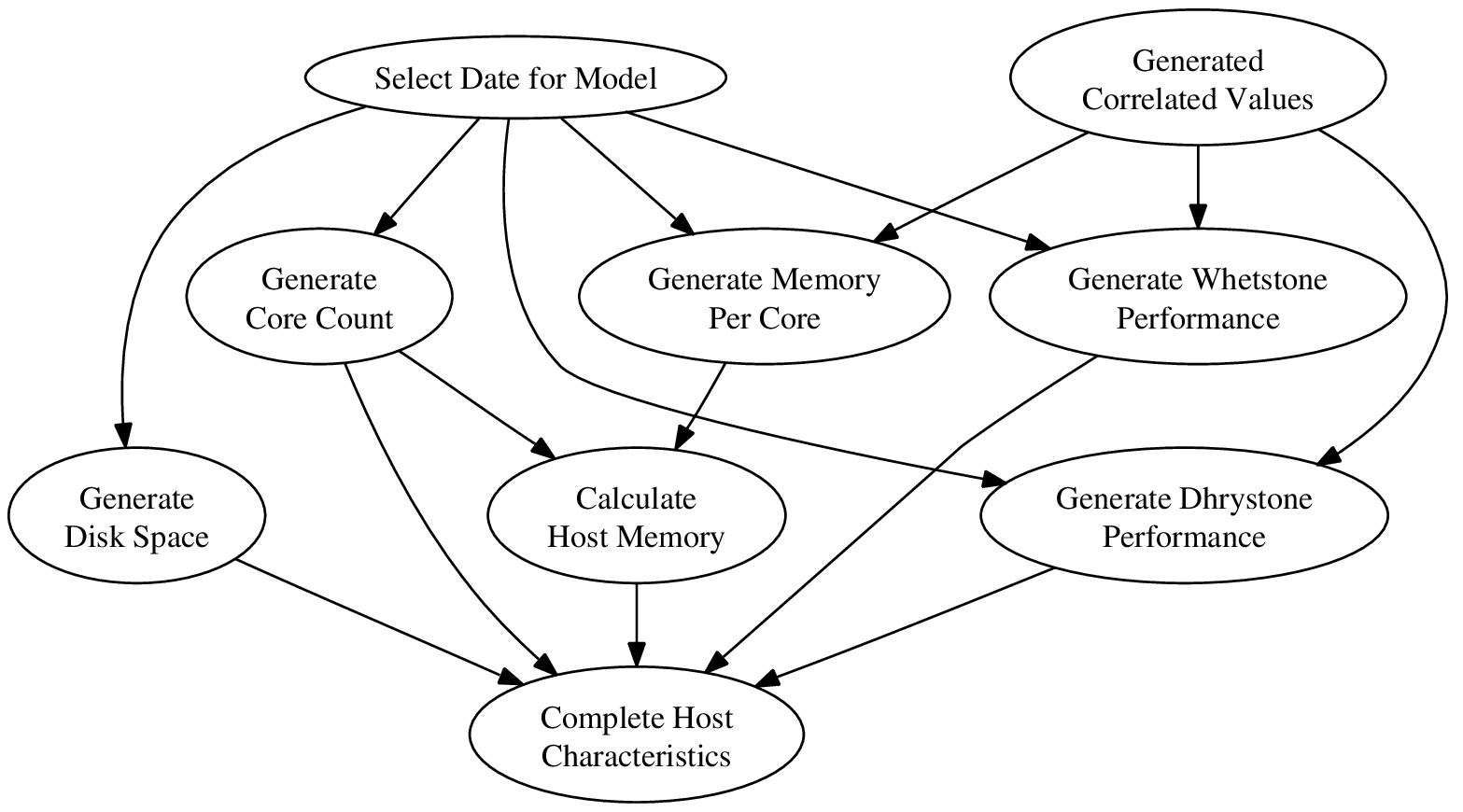}
\caption{Flowchart of host creation.}
\label{fig-host-creation}
\end{figure}

Next we use the model developed in the last section to generate hosts at a specified point in time.  We use standard statistical methods to validate the generated hosts and compare them to actual data.  Finally we use our model to offer predictions of how host composition will change up to the year 2014.

\subsection{Model Based Host Generation}
Figure \ref{fig-host-creation} shows the flowchart of host creation using our model.  First the user selects the date for the generated host.  Using the date, a core count is generated by using the ratios of cores to create a discrete probability distribution and selecting the number of cores with a uniform random number.

Using the method described in Section \ref{sec-model-speed}, correlated values are generated to create per-core-memory and processor benchmark speeds.  Similar to core count, the per-core-memory is selected using the ratio equations from Section \ref{sec-model-mem} to generate a discrete probability distribution which is then sampled.  Total memory is calculated by multiplying per-core-memory by the number of cores.  The benchmark values are generated by using the correlated normal values and re-normalizing them to the mean and variance predicted using values from Table \ref{benchmark-ratio-val-table}.  Available disk space is independent of other benchmarks, so it is generated by sampling a lognormal distribution with mean and variance predicted using values from Table \ref{benchmark-ratio-val-table}.

\subsection{Model Validation}

\begin{figure*}[!t]
\centering
\includegraphics[width=0.98\textwidth]{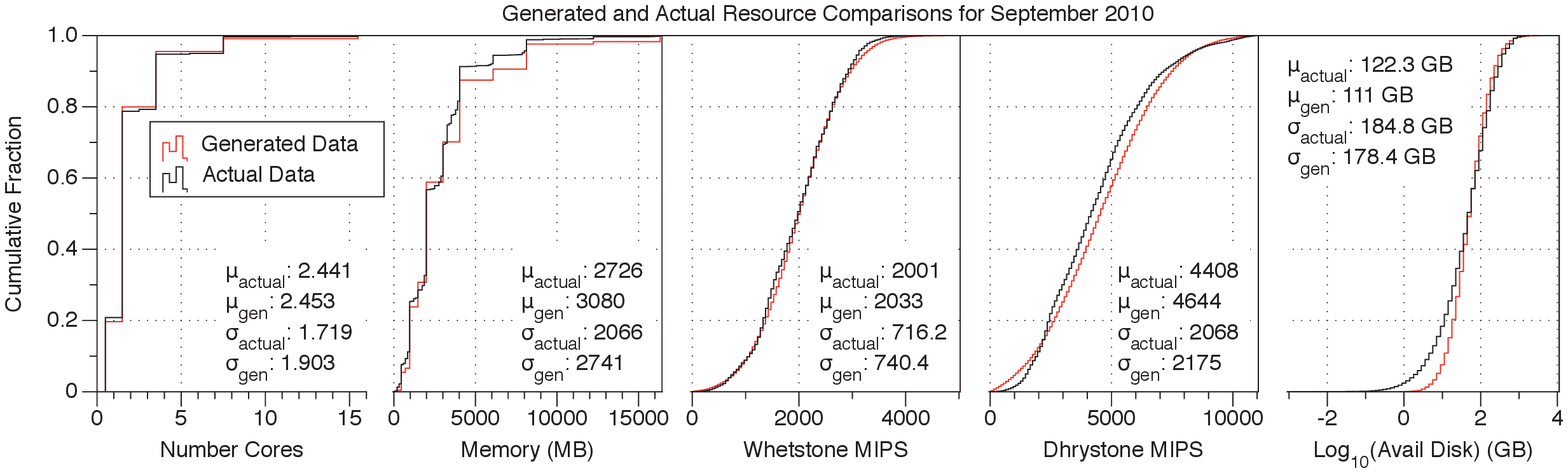}
\caption{Comparison of generated and actual data.}
\label{fig-gen-real-comparison}
\end{figure*}

\begin{table}
\caption{Correlation coefficients between generated hosts.}
\centering
\scriptsize
\begin{tabular}{|c|c|c|c|c|c|c|}
\hline
& Cores & Memory & Mem/Core & Whet & Dhry & Disk \\
\hline
Cores & 1.00 & \cellcolor[gray]{.7} \textbf{0.727} & 0.014 & 0.004 & 0.011 & -0.003 \\
\hline
Memory & \cellcolor[gray]{.7} \textbf{0.727} & 1.00 & \cellcolor[gray]{.8} \textbf{0.544} & 0.162 & 0.139 & -0.002 \\
\hline
Mem/Core & 0.014 & \cellcolor[gray]{.8} \textbf{0.544} & 1.00 & \cellcolor[gray]{.9} \textbf{0.307} & \cellcolor[gray]{.9} \textbf{0.251} & -0.002 \\
\hline
Whet & 0.004 & 0.162 & \cellcolor[gray]{.9} \textbf{0.307} & 1.00 & \cellcolor[gray]{.8} \textbf{0.505} & -0.002 \\
\hline
Dhry & 0.011 & 0.139 & \cellcolor[gray]{.9} \textbf{0.251} & \cellcolor[gray]{.8} \textbf{0.505} & 1.00 & -0.003 \\
\hline
Disk & -0.003 & -0.002 & -0.002 & -0.002 & -0.003 & 1.00 \\
\hline
\end{tabular}
\label{gen-corr-table}
\end{table}

Using our model in combination with this technique, we
generate a set of sample hosts for September 1, 2010.
Figure \ref{fig-gen-real-comparison} shows CDFs of the
generated and actual data for September 1, 2010.  The
generated values are close to the actual data, with means
ranging from a difference of 0.5\% for cores up to 13.0\%
for host memory and standard deviations ranging from a
difference of 3.5\% for Whetstone up to 32.7\% for memory.
We also generated QQ-plots for the data and visually confirmed
the fit of the generated hosts.  These plots are not included
in this paper for space reasons.

Table \ref{gen-corr-table} shows the correlation coefficients between hosts in the generated data for September 2010 calculated in the same way as Table \ref{corr-table}.  The correlation between cores and memory for generated hosts is $r \approx 0.7$ which matches the actual data $r \approx 0.6$.  This is promising for our model, since we do not explicitly correlate the random number generation for these resources.  Dhrystone and Whetstone benchmarks have a correlation of $r \approx 0.5$, also very close to the actual data correlation of $r \approx 0.6$.  The benchmarks also well match the per-core-memory correlation of $r \approx 0.3$.  Like the actual data, generated host disk space has almost no correlation.  The generated host memory is not as well correlated with the benchmarks ($r \approx 0.1$) as in the actual data ($r \approx 0.3$), but this correlation is not large so it should not greatly affect the generated model.

\subsection{Model Based Prediction}
\begin{figure}[!t]
\centering
\includegraphics[width=0.47\textwidth]{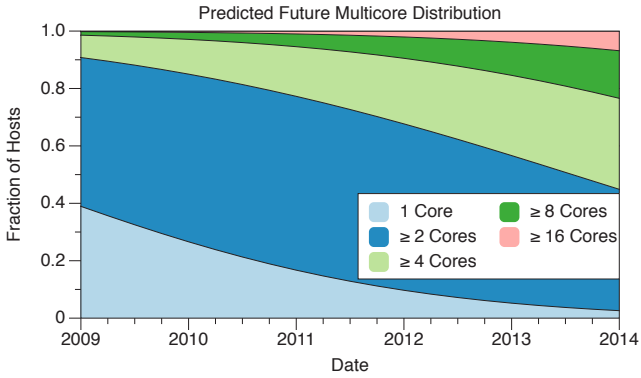}
\caption{Predicted fractions of host multicore CPUs.}
\label{fig-ncpu-prediction}
\end{figure}

Given the equations of resource ratios from Section \ref{sec-modelling} we can make predictions about how the host resource composition will change in the future.  Figure \ref{fig-ncpu-prediction} shows the predicted distribution of multicore processors over the next three years.  Based on the other equations, we estimate values of $a=12$, $b=-0.2$ to calculate the ratio of 8:16 cores.

There are several notable aspects of this prediction.  First, the number of single core hosts decreases to a negligible fraction within three years, as one would expect due to part failure and decreasing usefulness of the older single core machines.  Second, there are still a large number of 2 core hosts which comprise roughly 40\% of the total by 2014.  The average number of cores per host in 2014 is predicted to be 4.6 which is significantly higher than the number of 3.7 obtained by extrapolating the values of Figure \ref{fig-res-overview}.

\begin{figure}[!t]
\centering
\includegraphics[width=0.47\textwidth]{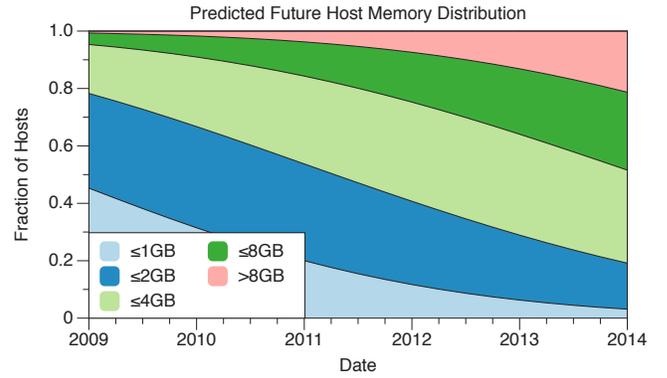}
\caption{Predicted fractions of hosts with specified total memory.}
\label{fig-mem-prediction}
\end{figure}

Figure \ref{fig-mem-prediction} shows the predicted distributions of total host memory over the next three years.  This prediction indicates an average of 6.8 GB per host by 2014 - very close to the value of 6.6 GB found by extrapolating values in Figure \ref{fig-res-overview}.  Using the values from Table \ref{benchmark-ratio-val-table} we predict the (mean, standard deviation) of Dhrystone as (8100, 4419), Whetstone as (2975, 868) and disk space as (272.0, 434.5) in 2014.

\ifthenelse{\boolean{long_version}}{

(**TODO) Best and worst hosts.  Given the developed model we can also make predictions about the best and worst hosts that will be available at a given time.
}{}

\section{Simulation Based Model Validation}
\label{sec-model-sim}
Finally, we perform simulations to demonstrate the value of
our model compared to other host resource representations.
Currently, most Internet-based computing applications have
focused on exclusively utilizing the CPU and most scheduling
algorithms aim to optimize the application makespan.
However, recent work has investigated using other resources,
such as disk space, to perform a wider range of services.
Certain applications may benefit disproportionally from
hosts with increased memory, greater processor speed or more
disk space.

Because of this, in these simulations we attempt to maximize total application utility of host resources rather than minimizing execution time.  Host utility can be thought of as how much benefit an application gets from running on a certain host.  We feel this is a better fit for analyzing our model since it includes all resource types and represents a generalized application that may desire a mix of resources or prefer certain resources over others.  To represent the utility of resources for a given application we use a variation on the well known Cobb-Douglas \cite{Douglas:1928p7797} utility function from economics.

Rather than the normal inputs of labor and capital, we use the resources for a host $H$: core count ($C_H$), memory ($M_H$), integer/floating point speed ($I_H$ and $F_H$) and disk space ($D_H$).  Then the utility $Y$ of running an application $A$ on host $H$ can be written as:
\begin{equation}
Y_A(H) = C_H^{\alpha} M_H^{\beta} I_H^{\gamma} F_H^{\delta} D_H^{\epsilon}
\end{equation}

where $\alpha$, $\beta$, $\gamma$, $\delta$, $\epsilon$ represent the utility returns to scale on each resource to the application.

\begin{table}
\caption{Simulation parameters for sample applications.}
\centering
\scriptsize
\begin{tabular}{|c|c|c|c|c|c|}
\hline
\multirow{3}{*}{Application} & Cores & Memory & Dhrystone & Whetstone & Disk \\
& ($\alpha$) & ($\beta$) & ($\gamma$) & ($\delta$) & ($\epsilon$) \\
\hline
SETI@home & 0.05 & 0.1 & 0.2 & 0.4 & 0.05 \\
\hline
Folding@home & 0.4 & 0.05 & 0.2 & 0.3 & 0.05 \\
\hline
Climate Prediction & 0.2 & 0.2 & 0.1 & 0.35 & 0.15 \\
\hline
P2P & 0.05 & 0.1 & 0.1 & 0.05 & 0.7 \\
\hline
\end{tabular}
\label{sample-app-table}
\end{table}

Table \ref{sample-app-table} shows the parameters we use for some sample applications in our simulation.  We chose these applications as a representative set of possible applications requiring Internet end hosts.  SETI@home represents an application doing radio signal analysis, which benefits from fast processing but does not require significant memory or disk space and does not utilize multiple cores.  Folding@home represents a parallel molecular dynamics simulation, which can use multiple cores and requires a medium amount of memory, but little disk.  Climate prediction requires a mix of all resources, with some emphasis on floating point speed.  P2P uses Internet end machines to perform distributed file sharing and benefits greatly from large disks, but has little use for processors or memory.

The simulation calculates the utility of each application running on each resource, then assigns resources to applications in a greedy round-robin fashion.  In the simulations we compare our correlated host synthesis model with two others.  The first is a simple model which uses extrapolation of the values in Figure \ref{fig-res-overview} and samples resource values from uncorrelated normal distributions (log-normal for disk space).  The second is based on the Grid resource model by Kee et. al. \cite{Kee:2004p763}.  This model uses a log-normal distribution for processors, a time and processor dependent model of memory and an exponential growth model for disk space.  We assign processor speed using the same method as the normal distribution model, and we use the same estimated mean/variance as our correlated model for the Grid resource model parameters where appropriate.  To make the comparison fair, we also update this model with more recent values from our analysis and generate a mix of older/newer hosts based on average host lifetime.

\begin{figure}[!t]
\centering
\includegraphics[width=0.46\textwidth]{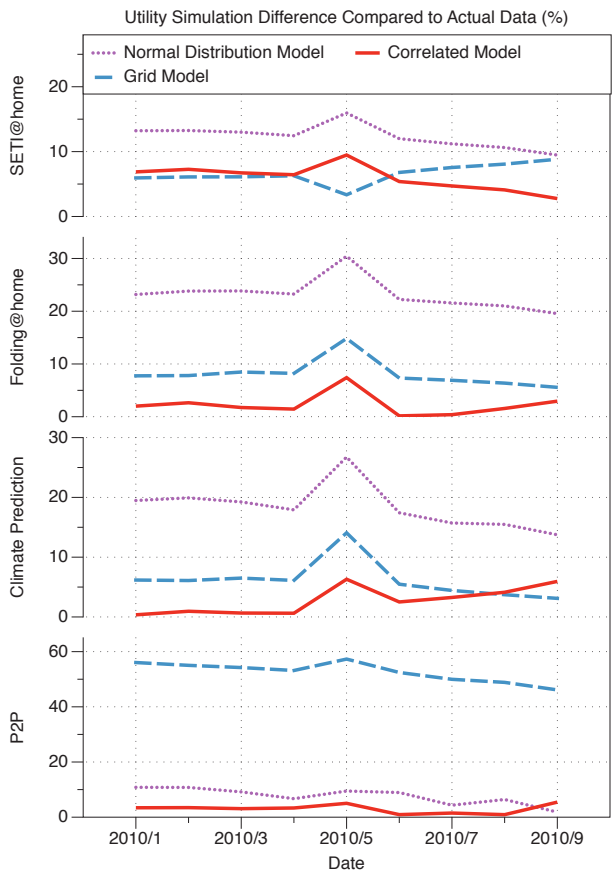}
\caption{Utility simulation results.}
\label{fig-sim-results}
\end{figure}

The simulation calculates the total utility for each application with the resources created by each model.  Figure \ref{fig-sim-results} shows the results for the simulation, comparing the normal distribution model, Grid resource model and correlated resource model described in this paper.  The simulations were run with data from January to September 2010.  The figure shows the percent difference between the total utility calculated using the specified model and the utility using the actual host data.  Multiple simulation runs showed little variance in results due to the large numbers of hosts involved.

The figure shows that the correlated model generally has a smaller difference with the actual data than the other models.  For the SETI@home application, the correlated model ranges between 3-10\% difference from the actual data, the Grid model between 3-9\% and the normal distribution model between 9-17\% difference.  The Folding@home application has a greater gap between the models, with the correlated model between 0-7\% difference, the Grid model between 5-15\% and the normal model around 20-31\% difference.  This is likely since the correlated model accurately captures the correlations between benchmark, memory and core count, which are all key components to the application.

The Climate Prediction application has similar results, with 0-7\% difference for the correlated model, 3-14\% difference for the Grid model and 14-28\% difference for the normal distribution model.  Again, the Climate Prediction application uses a mix of resources and will therefore be sensitive to the correlations between them.  The P2P application shows a major difference between the models, with a 0-5\% difference for the correlated model, 46-57\% difference for the Grid model and 0-11\% difference for the normal distribution model.  This is because the Grid model uses an exponential growth rule for disk space, which overestimates the available space.

Based on these results, we have shown that our model more closely reflects actual host resources, resource correlations and time dependent behavior.  Our model is significantly more accurate than simpler distribution models or other Grid models using uncorrelated distributions to model host resources.

\section{Conclusion}
\label{conc-sec}

Models of resources of Internet end hosts are critical for
the design and implementation of desktop software and
Internet-distributed applications.  We derive a model using
hardware traces of 2.7 million hosts on the Internet from
the SETI@home project. 

The following are our main contributions:

\begin{enumerate}
\item We determine a statistical model of the hardware
  resources of Internet hosts, namely, the number of cores,
  host memory, floating/integer speeds, and disk space (see
  Table~\ref{model-summary-table}).  This model captures:
 
  \begin{enumerate}
  \item the correlations among resources (in particular,
    between total memory and number of cores, or integer and
    floating point speeds)
  \item the evolution in time of resources (in particular,
    trends in the fraction of hosts with a certain number of
    cores or memory)
  \end{enumerate}
  
  Table \ref{model-summary-table} shows a condensed version
  of the model developed and evaluated in this paper.  This
  includes the resources described by the model, how they
  are derived and the $a$ and $b$ values used in the
  equation $a e^{b (year-2006)}$ describing either relative
  ratios or changes in the mean and variance of
  distributions.

\item We evaluate the accuracy in the
  context of a resource allocation problem for Internet-distributed applications.  Compared with naive models and
  Grid resource models, our model is up to 57\% more
  accurate.

\item Our resource trace data, and tools for automated model
  generation are available publicly at:
  
   \url{http://abenaki.imag.fr/resmodel/}
\end{enumerate}


There are several possible ways our model could be expanded.
First, the model of resources could be tied to models of
network topology and traffic, or models of host
availability, which would be useful for Internet-distributed
applications.  Second, the ideal distributions or resource correlations may
change over time, particularly for multiple cores, which could affect the model.
Finally, the use of GPUs for high performance computing is becoming common,
so with more data a GPU model could be developed as well.

\begin{table}
\caption{Summary of Model Parameters.}
\scriptsize
\centering
\begin{tabular}{|c|c|c|c|c|}
\hline
Resource & Value & Method & $a$ & $b$ \\
\hline
Cores & 1:2 Core & Relative Ratio & 3.369 & -0.5004 \\
\hline
& 2:4 Core & Relative Ratio & 17.49 & -0.3217 \\
\hline
& 4:8 Core & Relative Ratio & 12.8 & -0.2377 \\
\hline
Mem/Core & 256MB:512MB & Relative Ratio & 0.5829 & -0.2517 \\
\hline
& 512MB:768MB & Relative Ratio & 4.89 & -0.1292 \\
\hline
& 768MB:1GB & Relative Ratio & 0.3821 & -0.1709 \\
\hline
& 1GB:1.5GB & Relative Ratio & 3.98 & -0.1367 \\
\hline
& 1.5GB:2GB & Relative Ratio & 1.51 & -0.0925 \\
\hline
& 2GB:4GB & Relative Ratio & 4.951 & -0.1008 \\
\hline
Dhrystone & Mean (MIPS) & Normal Dist. & 2064 & 0.1709 \\
\hline
& Variance & Normal Dist. & 1.379e6 & 0.3313 \\
\hline
Whetstone & Mean (MIPS) & Normal Dist. & 1179 & 0.1157 \\
\hline
& Variance & Normal Dist. & 3.237e5 & 0.1057 \\
\hline
Disk Space & Mean (GB) & Lognorm Dist. & 31.59 & 0.2691 \\
\hline
& Variance & Lognorm Dist. & 2890 & 0.5224 \\
\hline
\end{tabular}
\label{model-summary-table}
\end{table}

\section*{Acknowledgements}
\label{Ack}
This work has been supported in part by the ANR project
Clouds@home (ANR-09-JCJC-0056-01).
\bibliographystyle{IEEEtran}

\bibliography{pubs,biblio}

\end{document}